\documentclass[floatfix,aps,showpacs,twocolumn,nofootinbib,amssymb,amsmath]{revtex4-1}

\usepackage{graphicx}
\usepackage{amsmath}
\usepackage{amsfonts}
\usepackage{amssymb}

\begin{document}

\title{Constraining Fundamental Physics with Future CMB Experiments}

\author{Silvia Galli$^{a,b}$, Matteo Martinelli$^{a}$ , Alessandro Melchiorri$^{a}$, Luca Pagano$^{a}$, Blake D. Sherwin$^{c}$, David N. Spergel$^{d}$}

\affiliation{$^a$ Dipartimento di Fisica and Sezione INFN,
Universit\`a di Roma ``La Sapienza'', Ple Aldo Moro 2, 00185, Rome, Italy}
\affiliation{$^b$ Laboratoire Astroparticule et Cosmologie (APC), Universit\'e Paris Diderot, 75205 Paris cedex 13, France}
\affiliation{$^c$ Department of Physics, Princeton University, Princeton, NJ 08544-1001, USA}
\affiliation{$^d$ Dept. of Astrophysical Sciences, Peyton Hall, Princeton University, Princeton, NJ 08544-1001, USA}

\begin{abstract}
The Planck experiment will soon provide a very accurate measurement of Cosmic Microwave Background anisotropies. This will let cosmologists determine most of the cosmological parameters with unprecedented accuracy. Future experiments will improve and complement the Planck data with better angular resolution and better polarization sensitivity. This unexplored region of the CMB power spectrum contains information on many parameters of interest, including neutrino mass, the number of relativistic particles at recombination, the primordial Helium abundance and the injection of additional ionizing photons by dark matter self-annihilation. We review the imprint of each parameter on the CMB and forecast the constraints achievable by future experiments by performing a Monte Carlo analysis on synthetic realizations of simulated data. We find that next generation satellite missions such as CMBPol could provide valuable constraints with a precision close to that expected in current and near future laboratory experiments. Finally, we discuss the implications of this intersection between cosmology and fundamental physics.
\end{abstract}


\maketitle

%%%%%%%%%%%%%%%%%%%%%%%%%%%%%%%%%%%%%%%%%%%%%%%%%%%%%%%%%%%%%%%%%%%%%%%%%
%%%%%%%%%%%%%%%%%%%%%%%%%%%%%%%%%%%%%%%%%%%%%%%%%%%%%%%%%%%%%%%%%%%%%%%%%
\section{Introduction}

%%%%%%%%%%%%%%%%%%%%%%%%%%%%%%%%%%%%%%%%%%%%%%%%%%%%%%%%%%%%%%%%%%%%%%%%%
%%%%%%%%%%%%%%%%%%%%%%%%%%%%%%%%%%%%%%%%%%%%%%%%%%%%%%%%%%%%%%%%%%%%%%%%%

Starting with COBE's groundbreaking detection of microwave background fluctuations \cite{cobe}, the past two decades have seen dramatic improvements 
in measurements of the microwave background temperature fluctuations (see e.g. \cite{altri} and \cite{wmap0}). Planck's highly anticipated temperature 
power spectrum measurements (see \cite{planck}) will further advance this program and produce significantly improved constraints on cosmological parameters. 

While Planck's measurement of the anisotropy power spectrum to multipoles $\ell \sim 2000$ will extract most of the information in primordial 
temperature fluctuations, ongoing and planned ground-based and balloon-based experiments are
exploring two important open frontiers: (a) the measurement of extremely ($\le5'$) small-scale temperature and polarization fluctuations \cite{small} 
and (b) the search for primordial B-modes, the distinctive signature of gravitational waves from inflation, on large scales \cite{bmodesgw}.

For example, balloon-borne experiments such as 
EBEX \cite{ebex} and SPIDER \cite{spider} will improve the measurements of CMB polarization 
while ground based telescopes such as
the Atacama Cosmology Telescope (ACT) \cite{ACTPol} and the South Pole Telescope \cite{spt}
will extend temperature and polarization measurements to smaller, sub-arcminute, angular scales.
Proposals for next generation CMB satellites such as CMBPol \cite{CMBPol} or B-POL \cite{bpol} are 
under evaluation from American and European space agencies.

What will we learn from measuring CMB temperature and polarization fluctuations on small-scales? The amplitude of temperature and 
polarization  fluctuations is determined by several different physical effects: (1) the amplitude of primordial fluctuations, 
(2) the evolution of the ionization fraction of the universe at $z > 1200$, which determines the sound speed for acoustic fluctuations, 
(3) the evolution of the ionization fraction of the universe at $z < 1200$, which determines the thickness of the surface of last scatter and 
(4) the transition from radiation to matter domination.
Moreover, while small-scale CMB fluctuations are initially pure $E$ mode, gravitational lensing rotates $E$ modes into $B$ modes \cite{cmblensing}. 
By measuring the pattern of small scale $E$ and $B$ modes, cosmologists will be able to determine the large-scale convergence field, 
a direct measure of the integrated density fluctuations between redshift $z=1100$ and $z=0$ (see e.g. \cite{hirata},\cite{okamotohu}). 
The convergence power spectrum is 
particularly sensitive to density fluctuations at $z \sim 2$, an important complement to planned optical lensing measurements 
that probe the evolution of density fluctuations in the $z < 1$ universe. 
%In this paper, we will focus on CMB-only constraints, 
%but anticipate that cross-correlations between CMB lensing and large-scale structure surveys will generate complementary measurements. HERE

The goal of this paper is to quantify the cosmological information 
that could come from these new datasets. This is important for several reasons. 
First, while there have been many studies of the future cosmological constraints from
Planck, very few papers have investigated the constraining power of
combinations of future CMB datasets from different sources. Second, as we will
describe in the next sections, we will consider a large set of parameters
focusing on those that mainly affect the "damping tail" of the CMB angular spectrum.
We consider additional parameters such as the total neutrino mass $\sum m_{\nu}$
(which affects the growth of structure in the late universe), 
the number of additional relativistic neutrino species
$N_{\nu}^{eff}$ (which changes the matter-radiation epoch), and possible changes in the recombination process due to
changes in the fractional helium abundance $Y_p$, dark matter self-annihilation
processes, and variations in fundamental constants such as as the fine structure constant $\alpha$ and 
Newton's gravitational constant $G$. We will not only show the constraints on each single
parameter but also the degeneracies among them.

We will consider $3$ experimental configurations: the Planck satellite \cite{planck},
the combination of Planck with ACT fitted with polarization-sensitive detectors, ACTPol, \cite{ACTPol} and, finally, the next
CMBPol satellite \cite{CMBPol}. 

Recent studies have already fully demonstrated the ability of next generation satellite missions to 
constrain inflationary parameters \cite{baumann} and the reionization history
\cite{cmbpolreio} in the framework of the CMBPol concept mission study (see also \cite{dode}).
For this reason we will not consider primordial gravitational waves, more general reionization 
scenarios or experiments that will mainly probe large angular scale polarization in this paper.

This paper will show that next generation CMB experiments can significantly improve constraints on cosmology and fundamental physics 
and could produce a detection of neutrino mass.
The paper is structured as follows. Section II describes our analysis approach. Section III presents our 
analysis for improved constraints from the planned ACTPol experiment and for the proposed CMBPol experiment. 
In Section IV we present our conclusions.

%%%%%%%%%%%%%%%%%%%%%%%%%%%%%%%%%%%%%%%%%%%%%%%%%%%%%%%%%%%%%%%%%%%%%%%%%
%%%%%%%%%%%%%%%%%%%%%%%%%%%%%%%%%%%%%%%%%%%%%%%%%%%%%%%%%%%%%%%%%%%%%%%%%
\section{Forecast Method and Assumptions}
%%%%%%%%%%%%%%%%%%%%%%%%%%%%%%%%%%%%%%%%%%%%%%%%%%%%%%%%%%%%%%%%%%%%%%%%%
%%%%%%%%%%%%%%%%%%%%%%%%%%%%%%%%%%%%%%%%

We generate synthetic datasets for the Planck, ACTPol and CMBPol experiments 
following the commonly used approach described 
for example in \cite{shimon} and \cite{lensextr}.
These datasets are generated starting from the assumption of a fiducial ``true'' cosmological model compatible with the WMAP-5 maximum likelihood parameters 
\cite{wmap5}, i.e. with baryon density $\Omega_{b}h^2=0.0227$, cold dark matter density 
$\Omega_{c}h^2= 0.110$, spectral index $n_s=0.963$, and optical depth $\tau=0.09$. 
This model also assumes a flat universe with a cosmological constant, 
massless neutrinos with effective number $N_{\nu}^{eff}=3.04$, standard recombination, Helium fraction $Y_p=0.24$ and
all fundamental constants fixed to their current values (we will vary all these parameters later).
Given the fiducial model, we use the publicly available 
Boltzmann code {\sc CAMB}\footnote{\tt
http://camb.info/}~\cite{Lewis:1999bs} to calculate the corresponding theoretical angular power
spectra $C_{\ell}^{TT}$, $C_{\ell}^{TE}$, $C_{\ell}^{EE}$ 
for temperature, cross temperature-polarization and  polarization.
\footnote{Note that we don't consider 
the $B$ mode lensing channel, we will discuss this choice later in this section}.

The synthetic datasets are then generated by considering for each $C_\ell$ 
a noise spectrum given by:

\begin{equation}
N_\ell = w^{-1}\exp(\ell(\ell+1)8\ln2/\theta^2),
\end{equation}

\noindent where $\theta$ is the FWHM of the beam assuming a Gaussian profile
and where $w^{-1}$ is the experimental power noise related to the 
detectors sensitivity $\sigma$ by $w^{-1} = (\theta\sigma)^2$.

We assume that beam uncertainties are small and that uncertainties due to foreground
removal are smaller than statistical errors.  These are demanding assumptions; however, the experimental groups are working hard to achieve these goals.

Together with the primary anisotropy signal we also take into account information
from CMB weak lensing, considering the power spectrum of the deflection
field $C_{\ell}^{dd}$ and its cross correlation with temperature maps
$C_{\ell}^{Td}$. A large number of methods have been suggested 
for lensing extraction from CMB maps. All these methods
exploit the non-gaussian signal induced by lensing. 
Here we use the quadratic estimator method of Hu \& Okamoto~\cite{okamotohu}, that provides an algorithm for
estimating the corresponding noise spectrum $N_{\ell}^{dd}$ from
the observed CMB primary anisotropy and noise power spectra.

This method doesn't include the polarization $BB$ channel since is dominated 
by the lensing signal. The Planck experiment is not sensitive to 
the $BB$ lensing signal, while the CMBPol experiment and, possibly, ACTpol
could detect it. While algorithms are available that can in principle
include in the forecast the lensing $BB$ signal, here we take the conservative
approach to not include it. This leaves open the possibility 
to use this channel for further checks for foregrounds contamination and systematics.

We generate mock datasets with noise properties consistent respectively with the Planck mission (see \cite{planck}),
the ACT telescope \cite{ACTPol} and the future CMBPol experiment \cite{CMBPol}.
For the simulated Planck dataset we consider the detectors at $70, 100,$ and $143GHz$
 while for ACTPol we use the single $150GHz$ channel. For CMBPol we
also consider the single $150GHz$ channel. The experimental specifications are reported in Table~\ref{tab:exp}
where the sensitivity $\sigma$ is in units of $\Delta T/T$.

Once a mock dataset is produced we compare a generic theoretical model
through a likelihood ${\cal L}$ defined as 

\begin{equation}
 - 2 \ln {\cal L} = \sum_{l} (2l+1) f_{\rm sky} \left(
\frac{D}{|\bar{C}|} + \ln{\frac{|\bar{C}|}{|\hat{C}|}} - 3 \right),
\label{chieff}
\end{equation}

where $D$ is defined as
\begin{eqnarray}
D  &=&
\hat{C}_\ell^{TT}\bar{C}_\ell^{EE}\bar{C}_V^{dd} +
\bar{C}_\ell^{TT}\hat{C}_\ell^{EE}\bar{C}_\ell^{dd} +
\bar{C}_\ell^{TT}\bar{C}_\ell^{EE}\hat{C}_\ell^{dd} \nonumber\\
&&- \bar{C}_\ell^{TE}\left(\bar{C}_\ell^{TE}\hat{C}_\ell^{dd} +
2\hat{C}_\ell^{TE}\bar{C}_\ell^{dd} \right) \nonumber\\
&&- \bar{C}_\ell^{Td}\left(\bar{C}_\ell^{Td}\hat{C}_\ell^{EE} +
2\hat{C}_\ell^{Td}\bar{C}_\ell^{EE} \right),
\end{eqnarray}

\noindent where $\bar{C}_l$ and $\hat{C}_l$ are the fiducial
and theoretical spectra plus noise respectively, 
and $|\bar{C}|$, $|\hat{C}|$ denote the determinants of
the theoretical and observed data covariance matrices respectively,

\begin{eqnarray}
|\bar{C}| &=& \bar{C}_\ell^{TT}\bar{C}_\ell^{EE}\bar{C}_\ell^{dd} -
\left(\bar{C}_\ell^{TE}\right)^2\bar{C}_\ell^{dd} -
\left(\bar{C}_\ell^{Td}\right)^2\bar{C}_\ell^{EE} ~, \\
|\hat{C}| &=& \hat{C}_\ell^{TT}\hat{C}_\ell^{EE}\hat{C}_\ell^{dd} -
\left(\hat{C}_\ell^{TE}\right)^2\hat{C}_\ell^{dd} -
\left(\hat{C}_\ell^{Td}\right)^2\hat{C}_\ell^{EE}~.
\end{eqnarray}

\noindent and finally $f_{sky}$ is the sky fraction 
sampled by the experiment after foregrounds removal.

We derive constraints from simulated data using a modified version of 
the publicly available Markov Chain Monte Carlo
package \texttt{cosmomc} \cite{Lewis:2002ah} with a convergence
diagnostic based on the Gelman and Rubin statistic performed on $8$ chains.
We sample the following nine-dimensional set of cosmological
parameters, adopting flat priors on them: the physical baryon and Cold Dark Matter density fractions,
$\omega_b=\Omega_bh^2$ and $\omega_c=\Omega_ch^2$, the ratio of the sound horizon to the angular diameter
distance at decoupling, $\theta_S$, the scalar spectral index $n_S$, the overall normalization of the spectrum 
$A_s$ at $k=0.002$ Mpc$^{-1}$, the optical depth to reionization, $\tau$, the total mass of neutrinos, $\sum m_{\nu}$, 
the primordial helium abundance, $Y_p$, and the dark energy equation of state $w$.
We also consider parameters that can change the process of recombination:
the dark matter self-annihilation rate $p_{ann}$, a variation in the fine structure constant
$\alpha/\alpha_0$ and in Newton's constant $\lambda_G=G/G_0$, where $\alpha_0$ and
$G_0$ are the currently measured values.
For these latter parameters we choose to sample the Hubble constant $H_0$ instead of $\theta_S$ 
since these parameters are derived assuming standard recombination.
We also use a cosmic age top-hat prior with 10 Gyr$ \le t_0 \le$ 20 Gyr.
Furthermore, we consider adiabatic initial conditions and we impose flatness.\\

In what follows we will consider temperature and polarization power spectrum data up to $\ell_{max}=2500$, due to possible
unresolved foreground contamination at smaller angular scales and larger multipoles. 
Measurements of small-scale temperature fluctuations by ACT\cite{act2} and SPT\cite{spt2} confirm that extragalactic 
foregrounds will limit precision measurements of  primordial temperature fluctuations to $\ell < 2500$. Even if these foregrounds could be removed, 
kinetic Sunyaev-Zel'dovich (KSZ) fluctuations would provide a limiting 
source of confusion that will be difficult to model and impossible to remove as it has the same spectral shape as primordial fluctuations. 
Small-scale polarization measurements offer our best hope to probe the early universe on angular scales of $\ell= 2000 - 4000$: 
dusty galaxies are thought to be only $1$-$2\%$ polarized \cite{forpol}. 
We expect that secondary fluctuations should produce minimal polarization fluctuations and therefore that polarization will
provide an unbeatable test for systematics. We have checked that including only polarization data from $\ell =2500$ 
up to $\ell_{max}=3500$ does not significantly change the cosmological results, again suggesting the use of high $\ell$ polarized data
for checks for systematics.

\begin{table}[!htb]%\footnotesize
\begin{center}
\begin{tabular}{rcccc}
Experiment & Channel & FWHM & $\Delta T/T$ & $\Delta P/T$ \\
\hline
Planck & 70 & 14' & 4.7 & 6.7\\
$f_{sky}=0.85$& 100 & 10' & 2.5 & 4.0 \\
& 143 & 7.1'& 2.2 & 4.2\\
\hline
ACTPol & 150 & 1.4' & 14.6& 20.4 \\
$f_{sky}=0.19$ & & & &\\
\hline
CMBPol & 150 & 5.6' &0.037& 0.052 \\
$f_{sky}=0.72$ & & & &\\ \hline
\end{tabular}
\caption{Planck \cite{planck}, ACTPol\cite{ACTPol} and CMBPol\cite{CMBPol} experimental specifications.  Channel frequency is given
in GHz, FWHM in arcminutes and noise per pixel in $10^{-6}$ for the Stokes I, Q and U parameters. In the analysis, we assume that beam uncertainties and foreground uncertainties are smaller than the statistical errors in each of the experiments.}
\label{tab:exp}
\end{center}
\end{table}

%%%%%%%%%%%%%%%%%%%%%%%%%%%%%%%%%%%%%%%%%%%%%%%%%%%%%%%%%%%%%%%%%%%%%%%%%
%%%%%%%%%%%%%%%%%%%%%%%%%%%%%%%%%%%%%%%%%%%%%%%%%%%%%%%%%%%%%%%%%%%%%%%%%

\section{Results}

%%%%%%%%%%%%%%%%%%%%%%%%%%%%%%%%%%%%%%%%%%%%%%%%%%%%%%%%%%%%%%%%%%%%%%%%%
%%%%%%%%%%%%%%%%%%%%%%%%%%%%%%%%%%%%%%%%%%%%%%%%%%%%%%%%%%%%%%%%%%%%%%%%%

\subsection{Constraints on the ``standard'' $6$ parameters $ \rm \Lambda-CDM$ scenario}

\begin{table}[!htb]
\begin{center}
\begin{ruledtabular}
\begin{tabular}{r|c|c c|c c}
Parameter  & Planck& \multicolumn{2}{c|}{Planck+ACTPol}&\multicolumn{2}{c}{CMBPol}\\
uncertainty & & & \\
\hline
$\sigma(\Omega_b h^2)$ & 0.00013 & 0.000078 &(1.7)& 0.000034 &(3.8)\\ 
$\sigma(\Omega_c h^2)$ & 0.0010 & 0.00064 &(1.6)& 0.00027 &(3.7)\\ 
$\sigma(\theta_s)$ & 0.00026 & 0.00016 &(1.6)& 0.000052 &(5.0)\\ 
$\sigma(\tau)$ & 0.0042 & 0.0034 &(1.2)& 0.0022 &(1.9)\\ 
$\sigma(n_s)$ & 0.0031 & 0.0021 &(1.5)& 0.0014 &(2.2)\\ 
$\sigma(\log[10^{10} As])$ & 0.013 & 0.0086 &(1.5)& 0.0055 &(2.4)\\ 
$\sigma(H_0)$ &0.53 &0.30 &(1.8)&0.12 &(4.4)\\
\end{tabular}
\end{ruledtabular}
\end{center}
\caption{$68 \%$ c.l. errors on cosmological parameters from future surveys. A ``standard'', $6$ parameters 
$\Lambda$-CDM scenario is assumed. The numbers in brackets show the improvement factor $i=\sigma_{Planck}/\sigma$ respect to the 
Planck experiment.}
\label{tabstandard}
\end{table}

In Table \ref{tabstandard} we report the future constraints on the parameters of a
``minimal'' cosmological model. Together with the standard deviations on each parameter we also
report, for ACTPol and CMBPol, the improvement factor for each parameter 
defined as the ratio $\sigma_{Planck}/\sigma$
where $\sigma$ is the error from Planck+ACTPol or CMBPol and $\sigma_{Planck}$ is the constraint from Planck.

As we can see in the Table, the combination of  Planck with ACTPol will improve by a factor
$\sim 1.5$ the constraints on most of the parameters derived from Planck alone.
CMBPol will improve by a factor $\sim 4$  the constraints on the
baryon density, $H_0$ and $\theta_s$, while the constraints on parameters as 
$n_s$ and $\tau$ are improved by a factor $\sim 2$.

\subsection{Future Constraints on Neutrino Masses}

\begin{figure}[h!]
\centering
 \includegraphics[angle=0,width=0.4\textwidth]{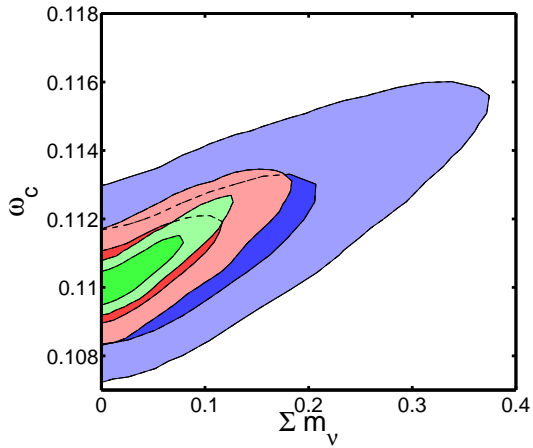}
\caption{68\% and 95\% likelihood contour plots on the $\sum m_\nu$ - $\omega_c$
 plane for Planck (blue), Planck+ACTPol (red) and CMBPol (green).}
\label{mnu_omegac}
\end{figure}

\begin{figure}[h!]
\centering
 \includegraphics[angle=0,width=0.4\textwidth]{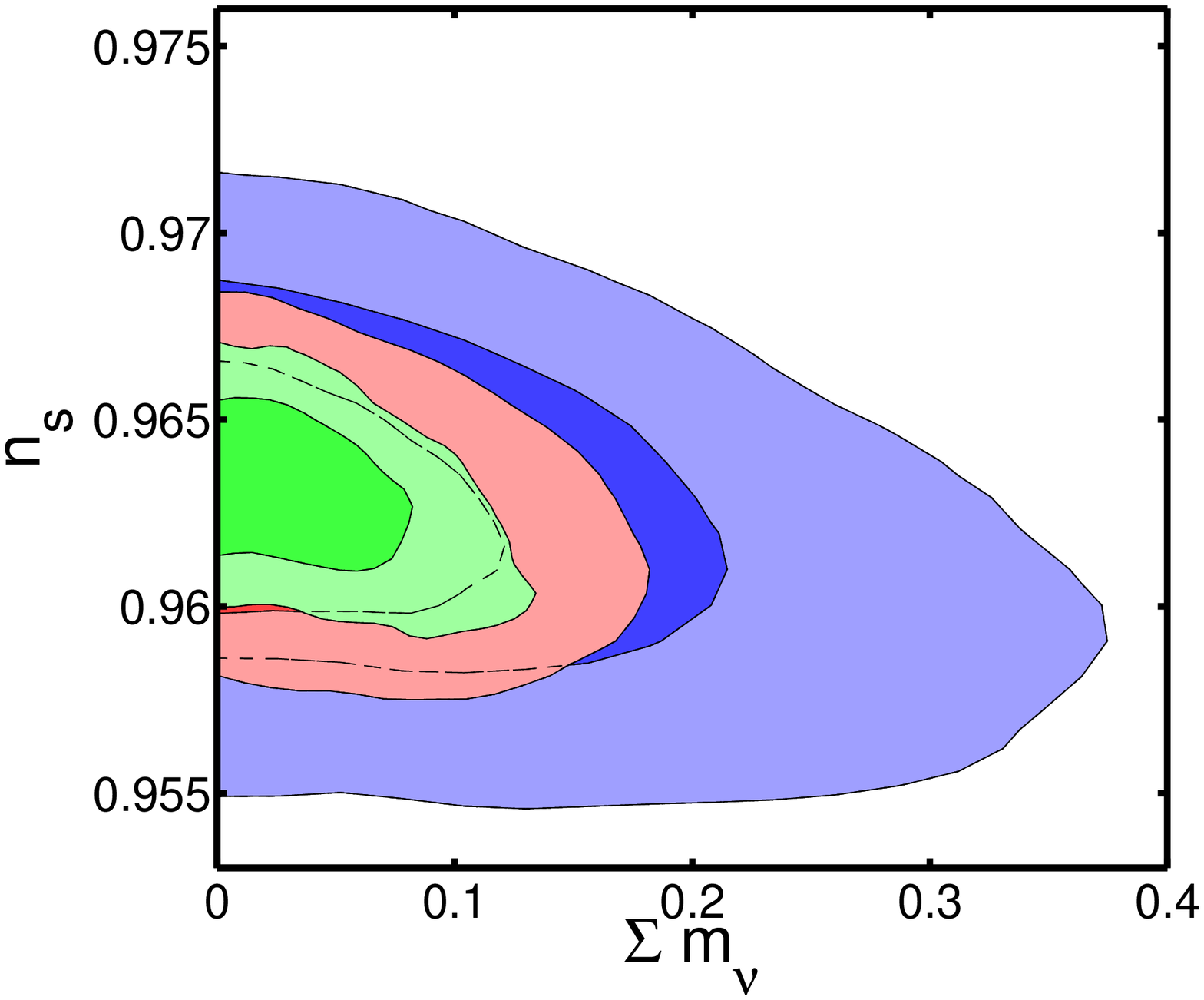}
\caption{68\% and 95\% likelihood contour plots on the $\sum m_\nu$ - $n_s$
 plane for Planck (blue), Planck+ACTPol (red) and CMBPol (green).}
\label{mnu_ns}
\end{figure}

\begin{table}[!htb]%\footnotesize
\begin{center}
\begin{ruledtabular}
\begin{tabular}{r|c|c c|c c}
Parameter& Planck&\multicolumn{2}{c|}{Planck+ACTPol}&\multicolumn{2}{c}{CMBPol} \\
uncertainty& & & \\
\hline
$\sigma(\Omega_bh^2)$ & $ 0.00014$ & $ 0.000081\ $&$(1.7) $&$0.000033\ $&$(4.2) $ \\
$\sigma(\Omega_ch^2)$ & $ 0.0017$  & $ 0.0010\ $&$(1.7) $&$0.00071\ $&$(2.4) $\\
$\sigma(\theta_S)$    & $0.00028$  & $0.00016\ $&$(1.7) $&$0.000062\ $&$(4.5) $\\
$\sigma(\tau)$        & $ 0.0042$ & $ 0.0034\ $&$(1.2) $&$0.0023\ $&$(1.8) $\\
$\sigma(n_S)$         &$0.0034$ & $0.0022\ $&$(1.5)  $&$0.0016\ $&$(2.1) $\\
$\sigma(\log[10^{10} A_S])$  & $0.013 $& $0.0094\ $&$(1.4)  $&$0.0065\ $&$(2.0) $\\
$\sigma(\sum m_{\nu})$  & $<0.16$&  $<0.08\ $&$(2.0) $&$<0.05\ $&$(3.2) $
\end{tabular}
\end{ruledtabular}
\caption{$68 \%$ c.l. errors on cosmological parameters in the case of massive neutrinos. 
The numbers in brackets show the improvement factor $\sigma_{Planck}/\sigma$ respect to the 
Planck experiment.}
\label{tabneumass}
\end{center}
\end{table}

The detection of the absolute mass scale of the neutrino is one of the major goals of experimental
particle physics. However, cosmology could provide an earlier, albeit model-dependent, detection.
CMB power spectra are sensitive to a total variation in neutrino mass eigenstates 
$\Sigma m_{\nu}$ (see e.g. \cite{ma,ichikawa1}) but can't discriminate between the mass 
of a single neutrino flavour (see e.g. \cite{slosar}) because of degeneracies with other parameters. 
Inclusion of massive neutrinos increases the anisotropy
at small scales because the decreased perturbation growth contributes to the photon energy density fluctuation.
Moreover, gravitational lensing leads to smoothing of the acoustic
peaks and enhancement of power on the damping tail of the power
spectrum; the amount of lensing is also connected to
the neutrino mass (see e.g.\cite{cmbneu}).

Current oscillation experiments provide essentially two mass
differences for the neutrino mass eigenstates: $\Delta m^2_{solar}\sim 8 \times 10^{-5} eV^2$ 
and $\Delta m^2_{atm} \sim 2.5\times 10^{-3} eV^2$ (see e.g. \cite{fogli} and references therein). 
An inverted hierarchy in the neutrino mass eigenstates predicts a lower limit to the total neutrino
mass of about $\sum m_{\nu} \ge 0.10 eV$ while a direct hierarchy predicts
$\sum m_{\nu} \ge 0.05 eV$. The goal for CMB experiments is therefore to
have a sensitivity better than $\sum m_{\nu} \le 0.10 eV$ for possibly ruling out the 
inverted hierarchy and better than $\sum m_{\nu} \le 0.05 eV$ for a definitive 
detection of neutrino mass.
As we can see from Table \ref{tabneumass} the expected sensitivity from 
Planck and Planck+ACTPol is sufficient to find the neutrino mass 
in the inverted hierarchy case, while CMBPol could possibly also measure it in the direct hierarchy case. In particular, the combination of ACTPol data with Planck is expected
to improve the bound on the neutrino mass by a factor of $2$ while CMBPol can improve it
by a factor of more than $3$. 
These limits are far better than those expected from future laboratory experiments.
The expected upper limit expected from the KATRIN \cite{katrin} beta decay experiment is
$m_{\nu_e}<0.2 eV$ at $90 \%$ c.l., which roughly translates to an upper limit of
$\sum m_{\nu} < 0.48 eV$ at one standard deviation (see \cite{elgaroy}).
Planck and Planck+ACTPol will explore the same energy scale, providing a great opportunity for confirming
or anticipating a mass detection from KATRIN. 
Planck alone will also falsify or confirm the claim of detection of the absolute scale of
the neutrino mass from the Heidelberg-Moscow neutrinoless double beta decay experiment
with a effective electron neutrino mass in the range $0.2 eV < m_{\nu_e} < 0.6 eV$ at 
$99.73 \%$ c.l. \cite{klapdor}.

Future double beta decay experiments such as MARE \cite{mare} should 
sample mass scales of the order of $m_{\nu_e}\sim 0.2 eV$. These experiments, if 
combined with Planck and Planck+ACTPol constraints could provide extremely valuable 
information on neutrino physics. For example, a CMB detection of a neutrino mass not confirmed 
by double beta decay experiments would rule out neutrinos as majorana-like particles.

Including a neutrino mass in the determination of the 
cold dark matter density $\omega_c$ results in an uncertainty that is
nearly doubled with respect to the standard analysis, as we can see by comparing Table \ref{tabneumass} with Table
\ref{tabstandard}. Moreover, the constraints
on $n_s$ are also affected.
We show in Figure \ref{mnu_omegac} and Figure \ref{mnu_ns} the $2$-D likelihood contour
plots at $68 \%$ and $95 \%$ confidence level in the $\Sigma m_{\nu}$ vs $\omega_c$ and vs $n_s$
planes respectively. As we can see, a non negligible neutrino mass has positive correlation with 
higher values of the cold dark matter abundance and lower values of the scalar spectral index.

\subsection{Future Constraints on Extra Background of Relativistic Particles}

\begin{figure}[h!]
\centering
 \includegraphics[angle=0,width=0.4\textwidth]{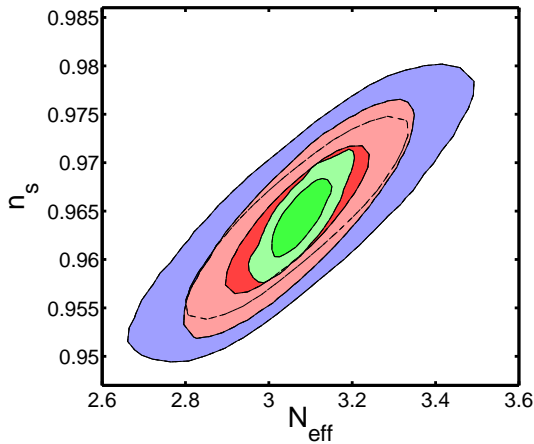}
\caption{68\% and 95\% likelihood contour plots on the $N_{eff} $ - $n_s$
 plane for Planck (blue), Planck+ACTPol (red) and CMBPol (green).}
\label{neff_ns}
\end{figure}

\begin{figure}[h!]
\centering
 \includegraphics[angle=0,width=0.4\textwidth]{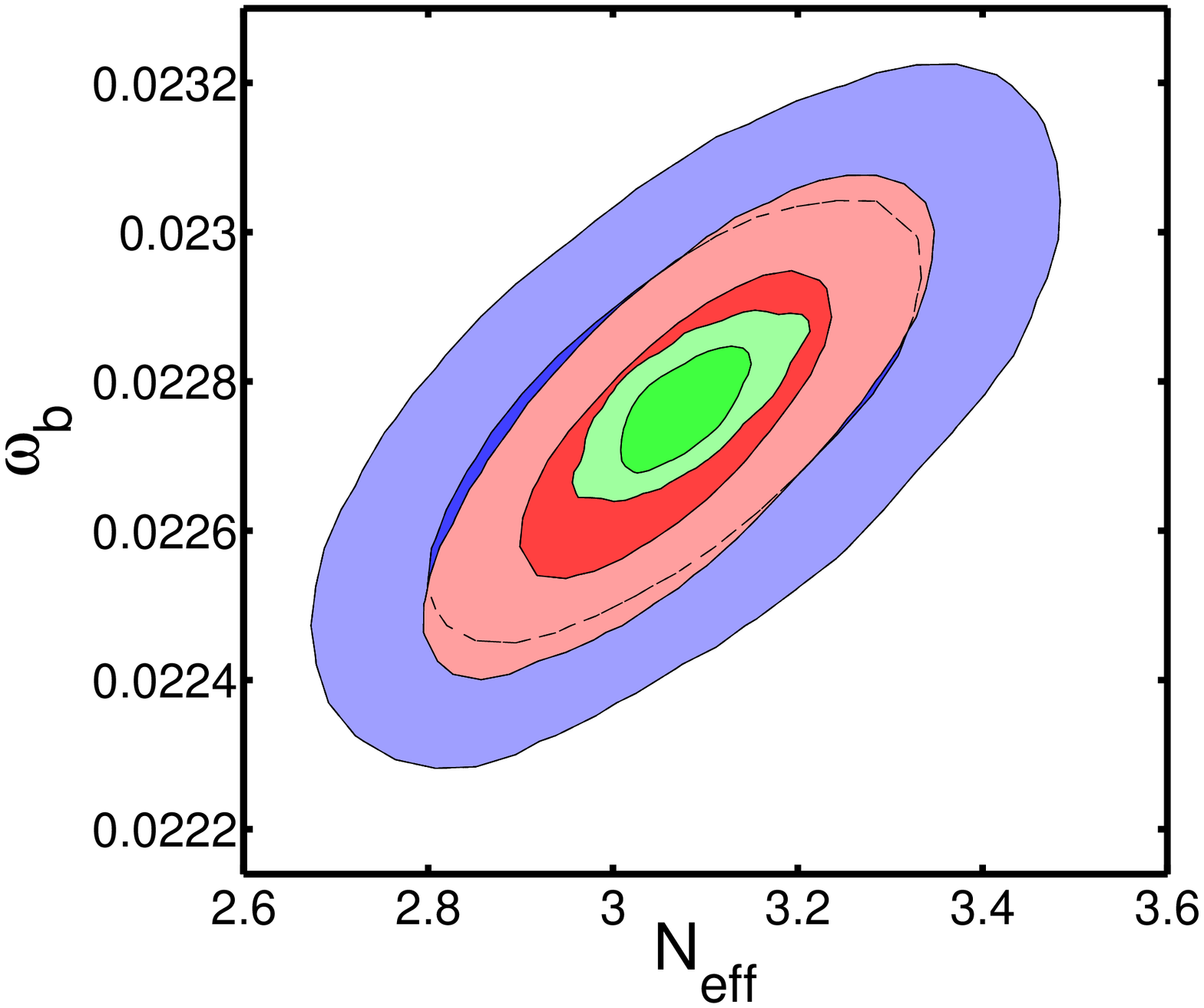}
\caption{68\% and 95\% likelihood contour plots on the $N_{eff} $ - $\omega_b$
 plane for Planck (blue), Planck+ACTPol (red) and CMBPol (green).}
\label{neff_omegab}
\end{figure}

\begin{figure}[h!]
\centering
 \includegraphics[angle=0,width=0.4\textwidth]{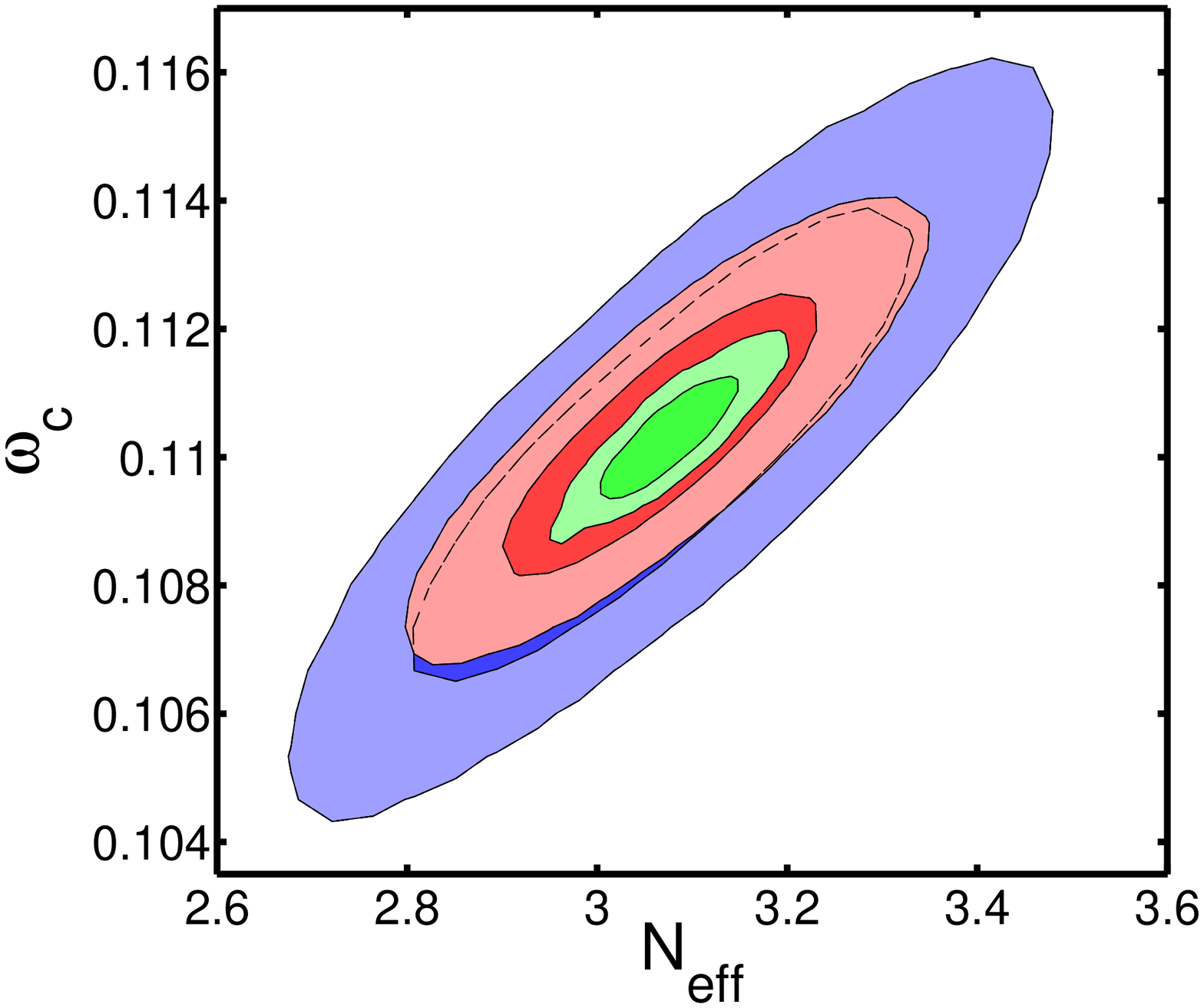}
\caption{68\% and 95\% likelihood contour plots on the $N_{eff} $ - $\omega_c$
 plane for Planck (blue), Planck+ACTPol (red) and CMBPol (green).}
\label{neff_omegac}
\end{figure}

\begin{figure}[h!]
\centering
 \includegraphics[angle=0,width=0.4\textwidth]{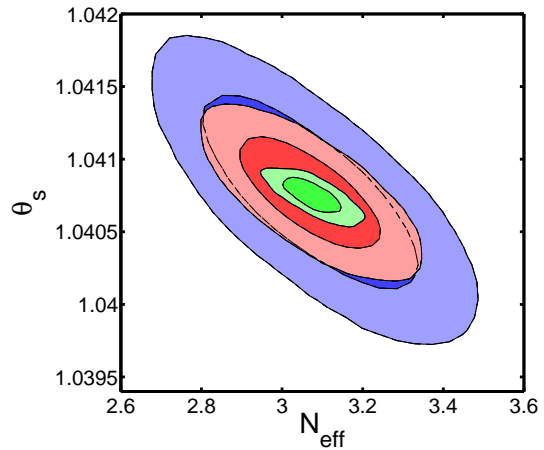}
\caption{68\% and 95\% likelihood contour plots on the $N_{eff} $ - $\theta_s$
 plane for Planck (blue), Planck+ACTPol (red) and CMBPol (green).}
\label{neff_thetas}
\end{figure}

An additional background of relativistic (and non-interacting) particles
can be parametrized by introducing an effective number of neutrino
species $N_{\nu}^{eff}$. This additional background
changes the CMB anisotropies through time variations of the
gravitational potential at recombination due to the presence of this 
non-negligible relativistic component (the so-called early Integrated Sachs Wolfe effect).
The main consequence is an increase in the small-scale CMB 
anisotropy (see e.g. \cite{bowen}). The results are reported in Table \ref{tabneurel}.

\begin{table}[h!tbp]
\begin{tabular}{r|c|c c|c c}
\hline\hline
Parameter& Planck&\multicolumn{2}{c|}{Planck+ACTPol}&\multicolumn{2}{c}{CMBPol} \\
uncertainty& & & \\
\hline
$\sigma(\Omega_b h^2)$ & 0.00020 & 0.00013 &(1.5) & 0.000048 &(4.1) \\ 
$\sigma(\Omega_c h^2)$ & 0.0025 & 0.0015 &(1.7) & 0.00058 &(4.3) \\ 
$\sigma(\theta_s)$ & 0.00044 & 0.00024 &(1.8) & 0.000075 &(5.9) \\ 
$\sigma(\tau)$ & 0.0043 & 0.0035 &(1.2) & 0.0023 &(1.9) \\ 
$\sigma(n_s)$ & 0.0073 & 0.0049 &(1.5) & 0.0026 &(2.8) \\ 
$\sigma(log[10^{10}A_s])$ & 0.019 & 0.013 &(1.5) & 0.0078 &(2.4) \\ 
$\sigma(N_{eff})$& 0.18 & 0.11 &(1.6) & 0.044 &(4.1) \\ \hline\hline
\end{tabular}
\caption{$68 \%$ c.l. errors on cosmological parameters in the case of extra background of relativistic
particles $N_{eff}$. The numbers in brackets show the improvement factor $\sigma_{Planck}/\sigma$ respect to the 
Planck experiment.}
\label{tabneurel}
\end{table}

As we can see, combining ACTPol with Planck will improve the constraint on
$N_{eff}$ by a factor of $1.5$ while CMBPol could improve it by a factor of $\sim 3.7$.
Comparing with the results in Table \ref{tabstandard}, the inclusion
of a background of relativistic particles strongly weakens the constraints on
$n_s$, $\omega_b$, $\omega_c$ and $\theta_s$. As we can see from Figures
\ref{neff_ns}, \ref{neff_omegab}, and \ref{neff_omegac} and \ref{neff_thetas}
there is indeed a strong positive correlation between $N_{\nu}^{eff}$ and these parameters
(negative for $\theta_s$).

While adding ACT data will improve the constraints by a factor $\sim 2$, CMBPol can provide
constraints that could give valuable information on the physics of neutrino 
decoupling from the photon-baryon primordial plasma. As it is well known,
the standard value of neutrino parameters $N_{eff}=3$ should be increased to
$N_{eff}=3.04$ due to an additional contribution from a partial heating of neutrinos during 
the electron-positron annihilations (see e.g. \cite{mangano1}). This effect, expected from standard physics, could be
tested by the CMBPol experiment, albeit at just one standard deviation.
However, the presence of non standard neutrino-electron interactions (NSI) may enhance the entropy transfer from electron-positron 
pairs into neutrinos instead of photons, up to a value of $N_{eff}=3.12$ (\cite{mangano2}). 
This value could be discriminated by CMBPol from $N_{eff}=3$ at $\sim 3$ standard deviations, shedding new light
on NSI models.

\subsection{Future Constraints on Dark Matter Self Annihilation}

Annihilating particles affect the ionization history of the universe in three
different ways: the interaction of the shower produced by the annihilation with the thermal gas can
 ionize the gas, induce Ly--$\alpha$ excitation of the hydrogen and heat
the plasma. The first two  modify the evolution of the free electron fraction $x_e$,
the third affects the temperature of baryons (\cite{dmannih},\cite{zhang},\cite{chenkam}). 

The rate of energy release $\frac{dE}{dt}$ per unit volume
by a relic self-annihilating dark matter particle is given by
\begin{eqnarray}
\label{enrateselfDM}
\frac{dE}{dt}(z)&=& \rho^2_c c^2 \Omega^2_{DM} (1+z)^6 p_{ann} \\
p_{ann}&=&f  \frac{<\sigma v>}{m_\chi}
\end{eqnarray}

where $n_{\rm DM}(z)$ is the relic DM abundance at a given redshift $z$,
$<\sigma v>$ is the effective self-annihilation rate and $m_\chi$ the mass
of our dark matter particle, $\Omega_{\rm DM}$ is the dark matter density
parameter and $\rho_c$ is the critical density of the universe today;
the parameter $f$ indicates the fraction of energy which is absorbed
{\it overall} by the gas, under the approximation that the energy absorption
takes place locally. The CMB is sensitive to the combined parameter $p_{ann}$ only. 
The greater $p_{ann}$, the higher the fraction of free electrons surviving after recombination, 
which widens the peak of the visibility function and dampens the peaks 
of the temperature and polarization angular power spectra.

\begin{figure}[h!]
\centering
\includegraphics[angle=0,width=0.5\textwidth]{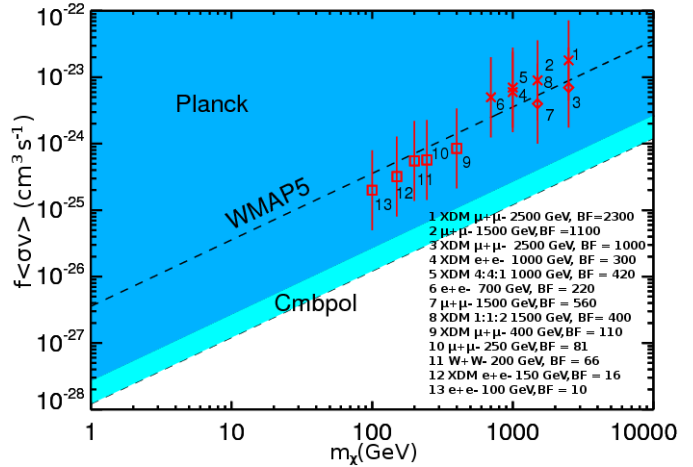}
\caption{Constraints on the self-annihilation cross-section at
recombination ($<\sigma v>)_{zr}$ times the gas-shower coupling parameter f . The dark blue area is excluded by Planck
 at 95\% c.l., whereas the lightest blue area indicates the additional parameter space excluded by CMBpol. Plank+ACT constraint is not shown as it is only 20\% tighter than the Planck constraint. The dashed line represents the current constraints given by the WMAP5 data \cite{dmannih}. The red data points are taken from \cite{slatyer} (based on the results of \cite{cholis} and \cite{grajek}),  and indicate the positions of models of dark matter particles that fit the observed cosmic-ray excesses for PAMELA data (squares), PAMELA+FERMI (diamods) and PAMELA+ATIC (crosses).
The ratios appearing in the legend indicate models of particles that annihilate through an intermediate light state to electrons, muons and pions
in the given ratio. Error bars indicate astrophysical uncertainties in the cross-section boost factor. We refer to \cite{slatyer} for further details on these models.}
\label{slatyer}
\end{figure}

\begin{table}[!h]%\footnotesize
\begin{center}
\begin{ruledtabular}
\begin{tabular}{r|c|c c|c l}
Parameter& Planck&\multicolumn{2}{c|}{Planck+ACTPol}&\multicolumn{2}{c}{CMBPol} \\
uncertainty& & & \\
\hline
$\sigma(\Omega_bh^2)$ & $ 0.00013 $  & $ 0.000079 $&$(1.6) $&$0.000032 $&$(4.1)$\\
$\sigma(\Omega_ch^2)$ & $ 0.0010$  & $ 0.00063 $&$(1.6)$&$0.00027 $&$(3.7)$\\
$\sigma(H_0)$     & $0.52$ & $0.30 $&$(1.7)$&$0.12 $&$(4.3)$\\
$\sigma(\tau)$         & $ 0.0042$ & $ 0.0034$&$(1.2)$&$0.0023 $&$(1.8)$\\
$\sigma(n_S)$          & $0.0032 $ & $0.0021$&$(1.5) $&$0.0015  $&$(2.1)$\\
$\sigma(\log[10^{10} A_S])$  & $0.013 $ & $0.0085$&$(1.5) $&$0.0055  $&$(2.4)$\\
$\sigma(p_{ann})$  & $<1.5\cdot 10^{-7}$ & $<1.2\cdot 10^{-7} $&$(1.2)$&$<6.3\cdot  10^{-8}\ $&$(2.4)$
\end{tabular}
\end{ruledtabular}
\caption{$68 \%$ c.l. errors on cosmological parameters in the case of dark matter annihilation. The upper limits on $p_{ann}$ are at $95\%$ c.l.. The parameter $p_{ann}$ is measured in $\rm [m^3/s/Kg]$.
The numbers in brackets show the improvement factor $\sigma_{Planck}/\sigma$ respect to the 
Planck experiment.}
\label{tabpann}
\end{center}
\end{table}

As we can see by comparing the entries in Table \ref{tabpann} with the results in 
Table \ref{tabstandard}, the inclusion of dark matter self-annihilation doesn't
substantially affect the constraints on the other parameters.

As shown in Galli et al. \cite{dmannih}, WMAP5 data already puts interesting constraints on dark matter annihilation, namely $p_{ann}=2.4\times 10^{-6}$ $\rm[m^3/s/Kg]$ at 95\% c.l..
This result disfavours dark matter annihilation as the main cause of the anomalies in the cosmic ray positron to electron fraction measured by PAMELA \cite{pamela}  and in the energy spectrum of cosmic ray electrons measured by ATIC \cite{atic} and less evidently by FERMI \cite{fermi}.
Slatyer et al. \cite{slatyer} examined the constraining power of this result on WIMP-like dark matter models that fit the excesses in the data. In these models, particles annihilate in leptons and pions both directly and through a new GeV-scale state. They showed that most of these models are excluded by WMAP5 at almost $2-\sigma$ c.l.. 

 Results reported in Table \ref{tabpann} will exclude these models at more than $\sim 10 - \sigma$ c.l.  for Planck and Plank+ACT and at $\sim 20 - \sigma$ for CMBPol, as shown in figure \ref{slatyer}. 

It is also interesting to notice that the constraints obtained by CMBpol are comparable to those obtained by a cosmic variance limited (CVL) experiment with angular resolution comparable to Planck and without lensing extraction. In fact, such a CVL experiment gives a constraint of $p_{ann}=5\times 10^{-8}$ \cite{dmannih}, comparable to the one reported in Table \ref{tabpann} for CMBpol.
Finally, it is worth noting that adding small scale data from ACT improves the constraints obtained with Planck only data by just 20\%. 
% \begin{table}[!htb]%\footnotesize
% \begin{center}
% \begin{tabular}{r|c|c|c|c}
% Parameter & Planck & Planck Conv & Planck+ACT &Planck+ACT Conv \\
% \hline
% $\Omega_bh^2$ & - & $ 0.00013 $ & - & $ 0.000079 $\\
% $\Omega_ch^2$ & - & $ 0.0010$ & - & $ 0.00063$\\
% $H_0$    & - & $0.52$& - & $0.00016$\\
% $\tau$        & - & $ 0.0042$& - & $ 0.0034$\\
% $n_S$         & - & $0.0032 $& - & $0.0021 $\\
% $\log[10^{10} A_S]$ & - & $0.013 $& - & $0.0085 $\\
% pann & - & $<1.5\cdot 10^{-7}$& - & $<3.6\cdot 10^{-8}$
% \end{tabular}
% \caption{1 sigma error. the upper limits on pann are at $95\%$ c.l.}
% \label{tab:results}
% \end{center}
% \end{table}

\subsection{Future Constraints on Helium Abundance}

\begin{figure}[h!]
\centering
 \includegraphics[angle=0,width=0.4\textwidth]{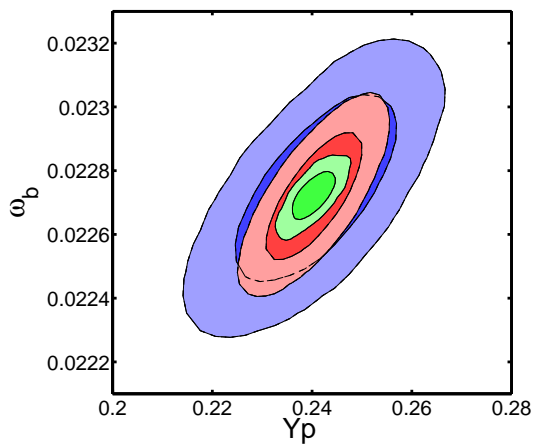}
\caption{68\% and 95\% likelihood contour plots on the $Y_{He}$ - $\omega_b$
 plane for Planck (blue), Planck+ACTPol (red) and CMBPol (green).}
\label{yhe_omegab}
\end{figure}

\begin{figure}[h!]
\centering
 \includegraphics[angle=0,width=0.4\textwidth]{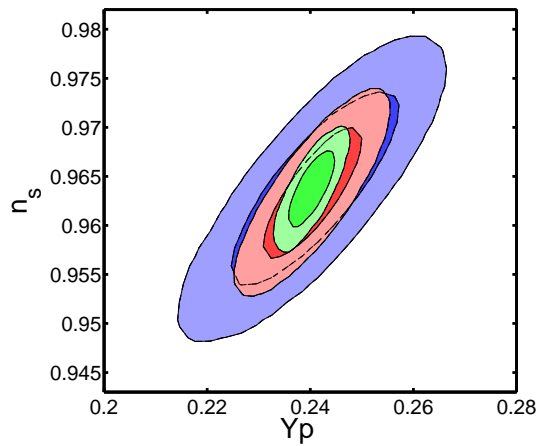}
\caption{68\% and 95\% likelihood contour plots on the $Y_{He}$ - $n_s$
 plane for Planck (blue), Planck+ACTPol (red) and CMBPol (green).}
\label{yhe_ns}
\end{figure}

As recently shown by several authors (\cite{trotta}, \cite{japan}, \cite{hamann}, \cite{komatsu7}) the small scale CMB
anisotropy spectrum can provide a powerful method for accurately determining the
primordial $\rm \phantom{}^4He$ abundance. Current astrophysical measurements of primordial fractional abundance~${ Y_p=^4He/(H+^4He)}$ can be contained in the conservative estimate of $Y_p=0.250 \pm 0.003$ (see e.g. \cite{iocco}).

As we can see from Table \ref{tabhe4}, the Planck satellite mission alone will
not reach this accuracy, even when combined with ACT. 
However it is important to note that the Helium abundance in the BBN scenario
is a growing function of $N_{eff}$ and the baryon density. A change in 
$\Delta N_{eff} \sim 1$ could produce a $\sim 5\%$ variation in $Y_p$ that could be measurable
by Planck or Planck+ACTPol. Moreover, a CMBPol-like experiment has
the potential of reaching a precision comparable with current astrophysical measurements.
This will open a new window of research for testing systematics in current 
primordial helium determinations.

\begin{table}[!htb]%\footnotesize
\begin{center}
\begin{ruledtabular}
\begin{tabular}{r|c|c c|c c}
Parameter& Planck&\multicolumn{2}{c|}{Planck+ACTPol}&\multicolumn{2}{c}{CMBPol} \\
uncertainty& & & \\
\hline
$\sigma(\Omega_bh^2)$  & $ 0.00019 $ &  $ 0.00013\ $&$(1.5) $&$0.000051\ $&$(3.7)$\\
$\sigma(\Omega_ch^2)$  & $ 0.0010$ &  $0.00064\ $&$(1.5)$&$0.00027\ $&$(3.7)$\\
$\sigma(\theta_S)$     & $0.00046$&  $0.00026\ $&$(1.8)$&$0.00010\ $&$(4.6)$\\
$\sigma(\tau)$         & $ 0.0043$&  $0.0035\ $&$(1.2)$&$0.0023\ $&$(1.9)$\\
$\sigma(n_S)$          & $0.0063 $&  $0.0042\ $&$(1.5)$&$0.0025\ $&$(2.5)$\\
$\sigma(\log[10^{10} A_S])$ & $0.013 $&  $0.013\ $&$(1.0) $&$0.0079\ $&$(1.6)$\\
$\sigma(Y_p)$  & $0.010$& $0.0060\ $&$(1.6)$&$0.0029\ $&$(3.4)$
\end{tabular}
\end{ruledtabular}
\caption{$68 \%$ c.l. errors on cosmological parameters in the case of helium abundance.
The numbers in brackets show the improvement factor $\sigma_{Planck}/\sigma$ respect to the 
Planck experiment.}
\label{tabhe4}
\end{center}
\end{table}

Comparing the results in Table \ref{tabhe4} with the constraints
obtained in the case of a standard analysis in Table \ref{tabstandard}, it is easy to see that
the major impact of including this parameter is on the determination
of the scalar spectral index $n_s$ and the baryon abundance, 
with the $1$-$\sigma$ c.l. increased by a factor $\sim 2$.
In Figures \ref{yhe_omegab} and \ref{yhe_ns} we 
plot the $2$-D likelihood contours at $68 \%$ and $95 \%$ c.l.
between $Y_p$ and these parameters.

\subsection{Future Constraints on Dark Energy Equation of State}
\begin{figure}[h!]
\centering
 \includegraphics[angle=0,width=0.4\textwidth]{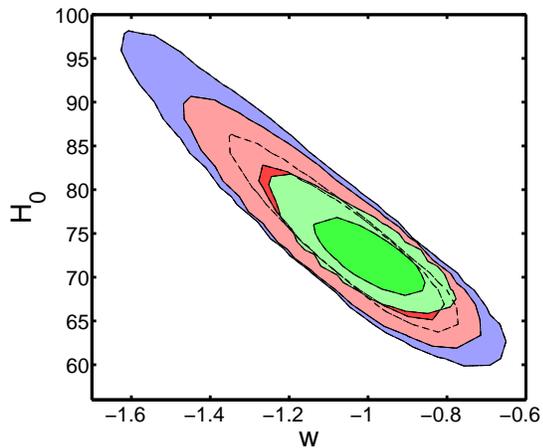}
\caption{68\% and 95\% likelihood contour plots on the $w$ - $H_0$
 plane for Planck (blue), Planck+ACTPol (red) and CMBPol (green).}
\label{w_h0}
\end{figure}

As is well known, primary CMB anisotropies are not able to provide accurate
measurements of the dark energy equation of state because of geometrical
degeneracies present with other parameters such as the amplitude of the 
dark energy density itself (see e.g. \cite{bean}) or the Hubble parameter.
However, the inclusion of CMB lensing can break these degeneracies (see
\cite{hulens},\cite{efsta}), as we can see in Table \ref{tabw} and from Figure \ref{w_h0}. 
It is interesting to notice that the error on w is strongly dominated by the degenaracy with $H_0$. 
In fact, the constraints on w from the 3 datasets considered are almost the same if one adds a strong 
prior on $H_0$ at a level of 2\%, obtaining $\sigma(w)=0.039$ for Planck, 
$\sigma(w)=0.037$ for Planck+ACT and $\sigma(w)=0.033$ for CMBpol. 

\begin{table}[htbp]
\begin{center}
\begin{ruledtabular}
\begin{tabular}{r|c|c c|c c}
Parameter& Planck&\multicolumn{2}{c|}{Planck+ACTPol}&\multicolumn{2}{c}{CMBPol} \\
uncertainty& & & \\
\hline
$\sigma(\Omega_b h^2)$ & 0.00013 & 0.000080 &(1.6)& 0.000032 &(4.2) \\ 
$\sigma(\Omega_c h^2)$ & 0.0011 & 0.00072 &(1.5) & 0.00038 &(3.0) \\ 
$\sigma(\theta_s)$ & 0.00026 & 0.00016 &(1.6) & 0.000053 &(4.9) \\ 
$\sigma(\tau)$ & 0.0040 & 0.0033 &(1.2) & 0.0023 &(1.8) \\ 
$\sigma(n_s)$ & 0.0032 & 0.0022 &(1.4) & 0.0016 &(2.0) \\ 
$\sigma(\log[10^{10} As])$ & 0.013 & 0.0098 &(1.3) & 0.0070 &(1.9) \\ 
$\sigma(w)$ & 0.2 & 0.15 &(1.3) & 0.085 &(2.4)\\ 
\end{tabular}
\end{ruledtabular}
\caption{$68 \%$ c.l. errors on cosmological parameters from future surveys in case of a variable dark energy equation of state $w$.
The numbers in brackets show the improvement factor $\sigma_{Planck}/\sigma$ respect to the 
Planck experiment.}
\label{tabw}
\end{center}
\end{table}

\subsection{Future Constraints on Variations of Fundamental Constants}
\begin{figure}[h!]
\centering
 \includegraphics[angle=0,width=0.4\textwidth]{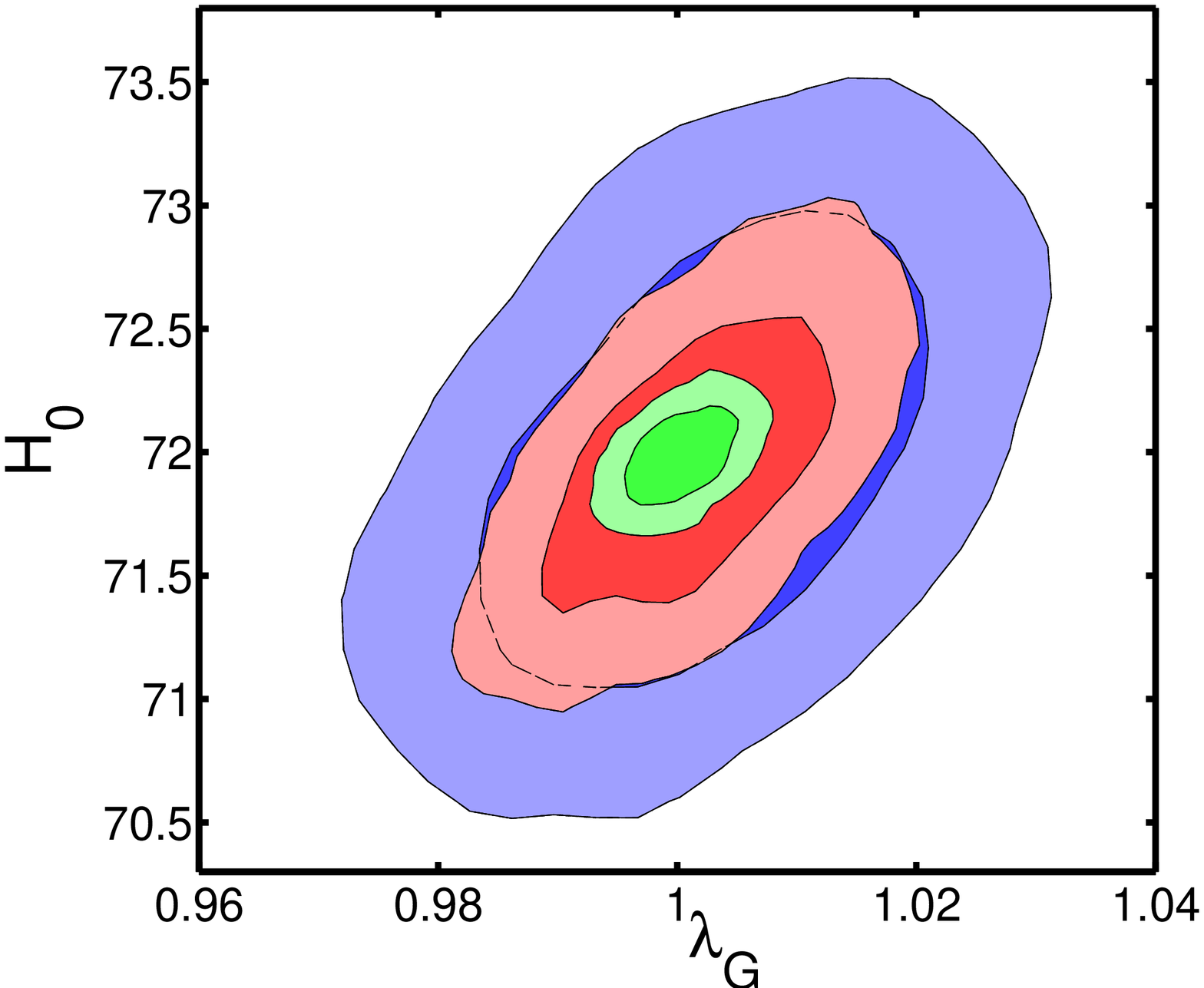}
\caption{68\% and 95\% likelihood contour plots on the $\lambda_G$ - $H_0$
 plane for Planck (blue), Planck+ACTPol (red) and CMBPol (green).}
\label{lambdag_h0}
\end{figure}
\begin{figure}[h!]
\centering
 \includegraphics[angle=0,width=0.4\textwidth]{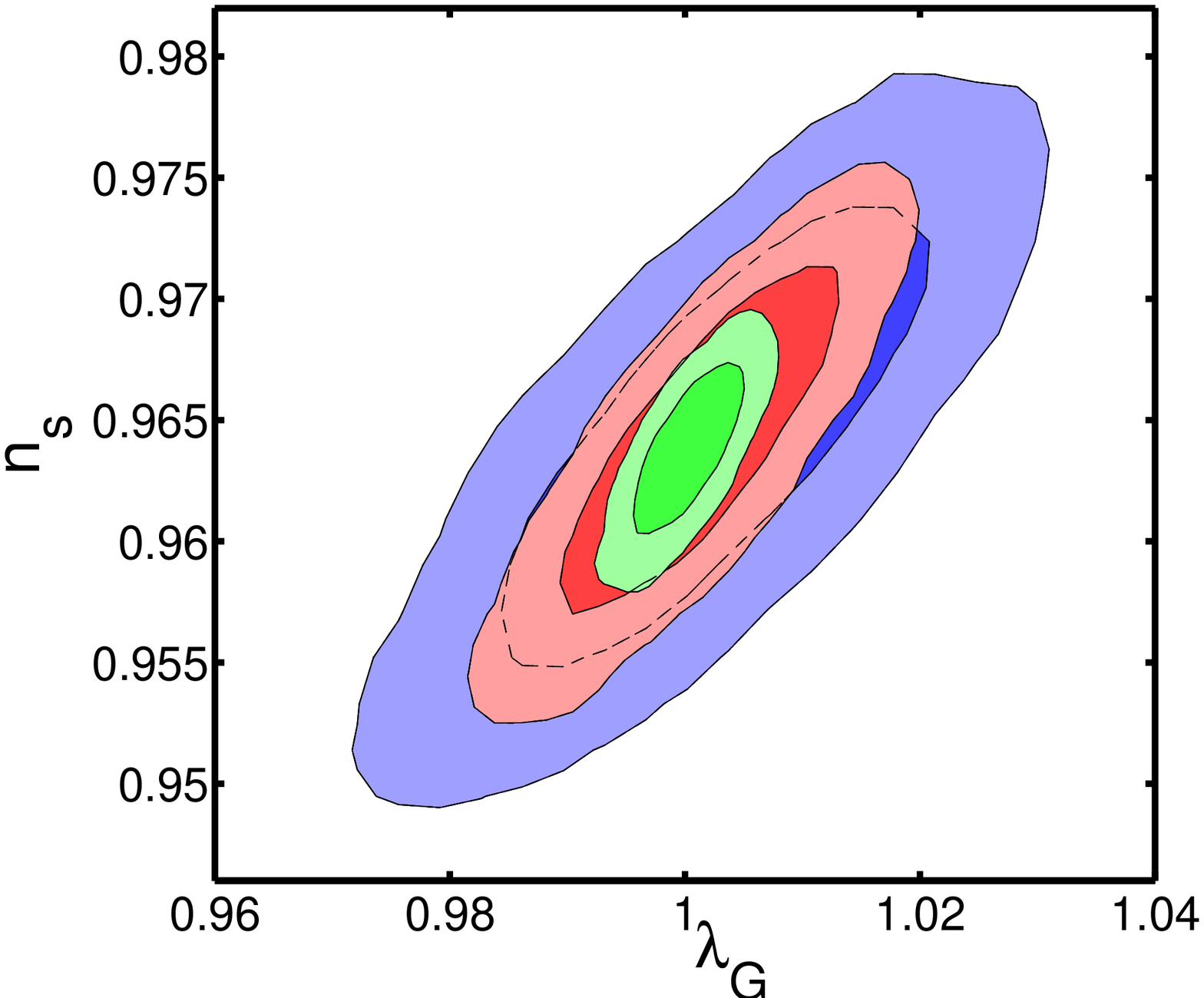}
\caption{68\% and 95\% likelihood contour plots on the $\lambda_G$ - $n_s$
 plane for Planck (blue), Planck+ACTPol (red) and CMBPol (green).}
\label{lambdaG_ns}
\end{figure}

\begin{figure}[htbp]
\centering
 \includegraphics[angle=0,width=0.4\textwidth]{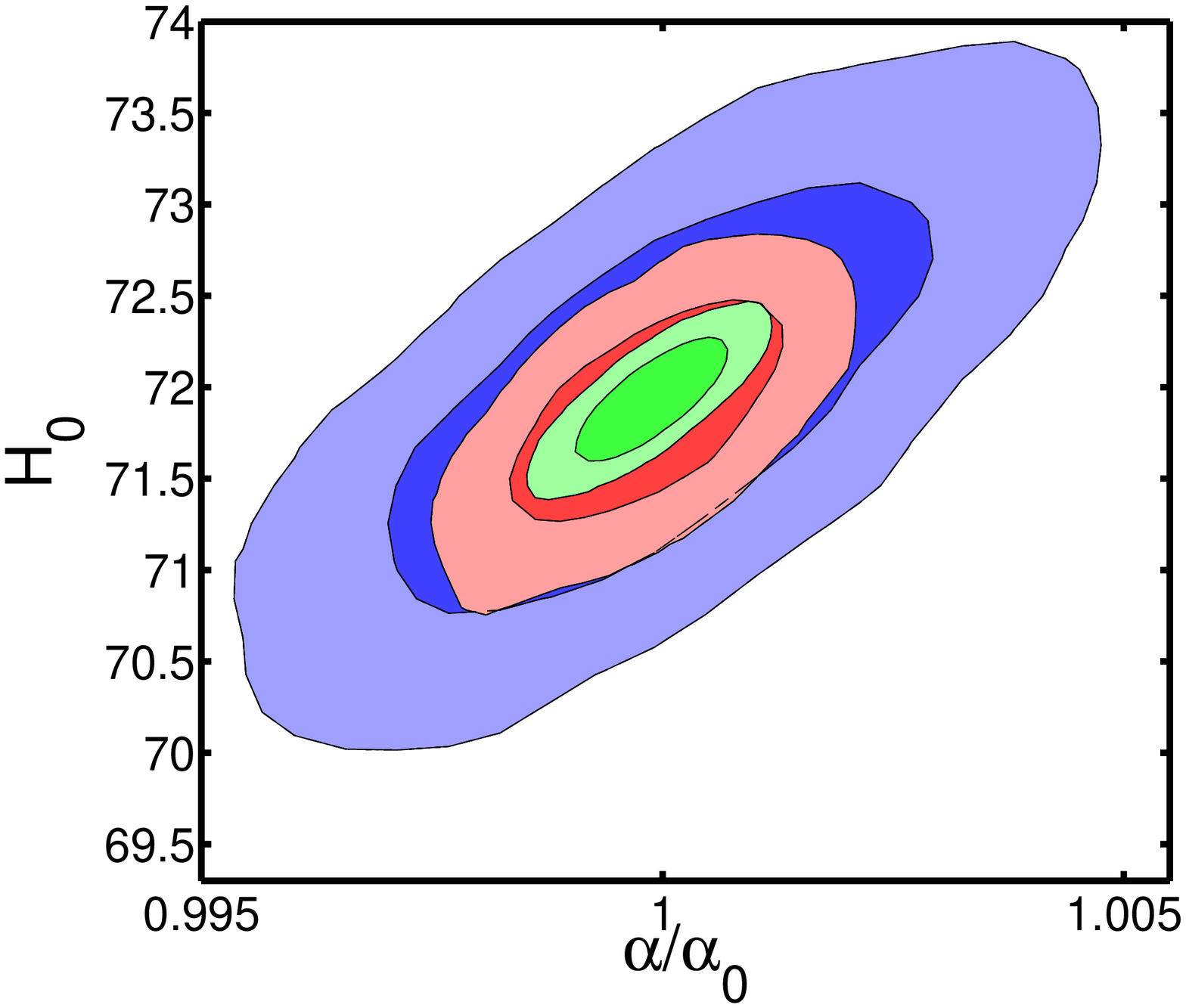}
\caption{68\% and 95\% likelihood contour plots on the $\alpha/\alpha_0$ - $H_0$
 plane for Planck (blue), Planck+ACTPol (red) and CMBPol (green).}
\label{alpha_h0}
\end{figure}
\begin{figure}[h!]
\centering
 \includegraphics[angle=0,width=0.4\textwidth]{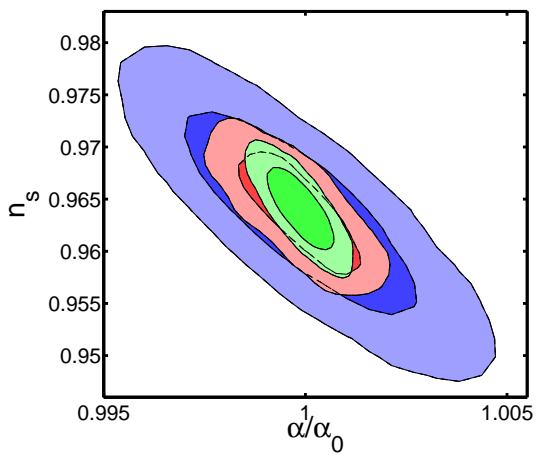}
\caption{68\% and 95\% likelihood contour plots on the $\alpha/\alpha_0$ - $n_s$
 plane for Planck (blue), Planck+ACTPol (red) and CMBPol (green).}
\label{alpha_ns}
\end{figure}

CMB anisotropies are sensitive to variations in fundamental constants
such as the fine structure constant $\alpha$ (see e.g. \cite{menegoni}, \cite{martins},\cite{ichikawa}, \cite{scoccola}) or Newton's constant $G$ (\cite{zahngalli}) through changes
in the recombination scenario. Varying $\alpha$ changes the ionization
and excitation rates and can delay or accelerate recombination. 
Varying $G$ does not affect recombination directly but "rescales" the 
expansion rate of the Universe, changing the epoch when recombination 
takes place.

The constraints are reported in Table \ref{tabalpha} and Table \ref{tabg}
for variations in $\alpha$ and $G$ respectively. In order to parametrize
the variations with dimensionless quantities we have considered variations
in the parameters $\Delta_{\alpha}=\alpha/\alpha_0$ e $\lambda_G=G/G_0$ where
$\alpha_0$ and $G_0$ are the current values of these fundamental constants,
measured in the laboratory\footnote{See http://www.codata.org/} $\alpha_0=7.297 352 5376(50) \times 10^{-3} $ 
and $G_0=6.674 28(67) \times 10^{-11} m^3 kg^{-1} s^{-2}$.

\begin{table}[h!tbp]
\begin{tabular}{r|c|c c|c c}
\hline\hline
Parameter& Planck&\multicolumn{2}{c|}{Planck+ACTPol}&\multicolumn{2}{c}{CMBPol} \\
uncertainty& & & \\
\hline
$\sigma(\Omega_b h^2)$ & 0.00019 & 0.00013 &(1.5)& 0.000048 &(3.9) \\ 
$\sigma(\Omega_c h^2)$ & 0.0010 & 0.00068 &(1.5) & 0.00025 &(4.0) \\ 
$\sigma(\tau)$ & 0.0042 & 0.0037 &(1.1) & 0.0022 &(1.9) \\ 
$\sigma(H_0)$ & 0.60 & 0.40 &(1.5) & 0.13 &(4.6) \\ 
$\sigma(n_s)$ & 0.0061 & 0.0046 &(1.3) & 0.0023 &(2.6) \\ 
$\sigma(\log[10^{10}A_s])$ & 0.018 & 0.013 &(1.4) & 0.0073 &(2.5) \\ 
$\sigma(\lambda_G)$ & 0.012 & 0.0076 &(1.6) & 0.0030 &(4.0) \\ \hline\hline
\end{tabular}
\caption{$68 \%$ c.l. errors on cosmological parameters from future surveys in case of a variable gravitational constant G.
The numbers in brackets show the improvement factor $\sigma_{Planck}/\sigma$ respect to the 
Planck experiment.}
\label{tabg}
\end{table}
\begin{table}[htbp]

\begin{tabular}{r|c|c c|c c}
\hline \hline
Parameter& Planck&\multicolumn{2}{c|}{Planck+ACTPol}&\multicolumn{2}{c}{CMBPol} \\
uncertainty& & & \\
\hline
$\sigma(\Omega_b h^2)$ & 0.00013& 0.000087 &(1.6) & 0.000035 &(4.1) \\ 
$\sigma(\Omega_c h^2)$ & 0.0012 & 0.00072 &(1.7) & 0.00032 &(3.9) \\ 
$\sigma(\tau)$ & 0.0042 & 0.0034 &(1.2) & 0.0024 &(1.8) \\ 
$\sigma(H_0)$ & 0.77 & 0.40 &(1.9) & 0.21 &(3.8) \\ 
$\sigma(n_s)$ & 0.0060 & 0.0036 &(1.8) & 0.0026 &(2.6) \\ 
$\sigma(\log[10^{10}A_s])$ & 0.015 & 0.011 &(1.4) & 0.0042 &(2.5) \\ 
$\sigma(\alpha/\alpha_0)$ & 0.0018 &0.00095  &(2.0) & 0.00053 &(3.7) \\ \hline\hline
\end{tabular}
\caption{$68 \%$ c.l. errors on cosmological parameters from future surveys in case of a variable fine structure constant $\alpha$.
The numbers in brackets show the improvement factor $\sigma_{Planck}/\sigma$ respect to the 
Planck experiment.}
\label{tabalpha}
\end{table}

As we can see from Tables \ref{tabalpha} and \ref{tabg}, a variation in these fundamental constants
has important effects for the determination of the scalar spectral index $n_s$ and the Hubble costant
$H_0$. This can also be seen in the $2$-D likelihood contour plots in Figures \ref{lambdag_h0}, 
\ref{lambdaG_ns}, \ref{alpha_h0}, and \ref{alpha_ns}.

\section{Conclusions}

In this paper we have performed a systematic analysis of the future constraints on several parameters achievable
from CMB experiments. Aside from the $5$ parameters of the standard
$\Lambda$-CDM model we have considered new parameters mostly related to quantities which
can be probed in a complementary way in the laboratory and/or with astrophysical measurements.
In particular we found that the Planck experiment will provide bounds on the sum of the masses $\Sigma m_{\nu}$ that 
could potentially definitively confirm or rule out the Heidelberg-Moscow claim for a detection of an absolute neutrino mass
scale. Planck+ACTPol could reach sufficient sensitivity for a robust detection of neutrino mass for an inverted hierarchy, while
CMBPol should also be able to detect it for a direct mass hierarchy.
The comparison of Planck+ACTPol constraints on baryon density, $N_{eff}$ and $Y_p$ with the
complementary bounds from BBN will provide a fundamental test for the whole cosmological scenario.
CMBPol could have a very important impact in understanding the epoch of neutrino decoupling. 
Moreover, the primordial Helium abundance can be constrained with an
accuracy equal to that of current astrophysical measurements but with much better control
of systematics.  Constraints on fundamental constants can be achieved at a level
close to laboratory constraints. Such overlap between cosmology and other fields of physics
and astronomy is one of the most interesting aspect of future CMB research.

We should note, however, that our forecasts rely on several technical assumptions. First, we assumed
that the theoretical model of the recombination process is accurately known. 
This is not quite true as corrections to the recombination process are already needed for the Planck experiment 
(see e.g. \cite{rubino}). However, this is mainly a computational problem that could be solved by the time of CMBPol launch, expected not before
2015. In addition, we assume that the foreground and beam uncertainties are smaller than the statistical errors.

Nevertheless, the results clearly show the advantage of adding small scale data from the ACT telescope
 to the Planck satellite data. Adding the former to the latter will improve the constraints by a factor $\sim2$ in most of the models considered.

\acknowledgments
This research has been supported by ASI contract I/016/07/0 ``COFIS.''
BS is supported by an NSF graduate fellowship. DNS
is supported by NASA theory Grant No. NNX087AH30G and NSF Grant No. 0707731.

%%%%%%%%%%%%%%%%%%%%%%%%%%%%%%%%%%%%%%%%%%%%%%%%%%%%%%%%%%%%%%%%%%%%%%%%%
%%%%%%%%%%%%%%%%%%%%%%%%%%%%%%%%%%%%%%%%%%%%%%%%%%%%%%%%%%%%%%%%%%%%%%%%%
\end{document}